\documentclass[%
 reprint,
nofootinbib,
 amsmath,amssymb,
 aps,
]{revtex4-2}

\usepackage[english]{babel}

\usepackage{latexsym}
\usepackage[T1]{fontenc}
\usepackage{amsmath}
\usepackage{amssymb}
\usepackage{hyperref}
\usepackage{cleveref}
\usepackage{enumerate}

\usepackage{graphicx}
\usepackage{dcolumn}
\usepackage{bm}

\def \limR {\lim_{R \ra \infty}}

\def \be {\begin{equation}}
\def \ee {\end{equation}}

\def \ume {{\scriptstyle{\frac{1}{2}}}}
\def \ra {\rightarrow}

\def \eqq {\equiv}

\def \eps {{\varepsilon}}

\def \A {{\cal A}}
\def \B {{\cal B}}
\def \C {{\cal C}}
\def \D {{\cal D}}
\def \F {{\cal F}}
\def \G {{\cal G}}

\def \L {{\cal L}}

\def \O {{\cal O}}

\def \V {{\cal V}}

\def \Z {{\cal Z}}

\def \Psio {{\Psi_0}}

\def \Rbf {{\bf R}}

\def \AO {{\cal A}({\cal O})}
\def \AO' {{\cal A}({\cal O}')}

\def \limR {\lim_{R \ra \infty}}

\newfam\Ssfam
\font\eleSs=cmss10 at12pt \font\sevenSs= cmss10 at 8pt \font\sixSs= cmss10 at 6pt

\textfont\Ssfam=\eleSs\scriptfont\Ssfam=\sevenSs%
\scriptscriptfont\Ssfam=\sixSs \def\Ss{\fam\Ssfam\eleSs}

\def\doppio#1{{\rm I}\kern-.1667em{\rm #1}}

\def\Q{\text{Q}\kern-.52em
    \text{\vrule height1.5ex width.5pt depth0pt}\kern.45em}

\def\Z{{\mathchoice {\hbox{$\Ss\textstyle Z\kern-0.4em Z$}}
{\hbox{$\Ss\textstyle Z\kern-0.4em Z$}} {\hbox{$\Ss\scriptstyle Z\kern-0.25em
Z$}} {\hbox{$\Ss\scriptscriptstyle Z\kern-0.2em Z$}}}}

\def\C{{\mathchoice{\hbox{$\rm\textstyle\text{\kern.35em\vrule
   height1.5ex width.5pt depth0pt\kern-.35em C}$}}
{\hbox{$\rm\textstyle\text{\kern.35em\vrule
   height1.5ex width.5pt depth0pt\kern-.35em C}$}}
{\hbox{$\rm\scriptstyle\text{\kern.35em\vrule
   height1.5ex width.3pt depth0pt\kern-.35em C}$}}
{\hbox{$\rm\scriptscriptstyle\text{\kern.35em\vrule
   height1.5ex width.2pt depth0pt\kern-.35em C}$}}}}

\def \be{\begin{equation} \displaystyle}
\def \ee{\end{equation}}

\def \A*{\mbox{$A^{*} $}}
\def \B*{\mbox{$B^{*} $}}
\def \C*{\mbox{$C^{*} $}}

\def \bea{\begin{eqnarray}}
\def \eea{\end{eqnarray}}

\def \be {\begin{equation} \displaystyle}

\def \ee {\end{equation}}

\def \ra {\rightarrow}
\def\AO {\mbox{${\cal A}({\cal O})$}}
\def\AO'{\mbox{${\cal A}({\cal O}')$}}
\def\O {\mbox{${\cal O}$}}

\def\A{\mbox{${\cal A}$}}

\def \ra{\rightarrow}

\def \eps {{\varepsilon}}

\def \O {{\cal O}}

\def \A {{\cal A}}
\def \AO {\A(\O)}
\def \AOl'{\A(\O_{loc}')}

\def \B {{\cal B}}
\def \E {{\cal E}}
\def \F {{\cal F}}
\def \D {{\cal D}}

\def \Fob  {\F_{obs}}
\def \LGF   {\L_{GF}}

\begin{document}

\preprint{APS/123-QED}

\title{A note on Gauss operators\\ and quantizations of Yang-Mills theories}

\author{Bruno Bucciotti}
\email{bruno.bucciotti@sns.it}
\affiliation{%
 Scuola Normale Superiore, Piazza dei Cavalieri 7, I-56126 Pisa, Italy\\
 INFN, Sezione di Pisa, Largo B. Pontecorvo, 3, I-56127 Pisa, Italy
}

\author{Franco Strocchi}%
 \email{franco.strocchi@sns.it}
\affiliation{%
 Dipartimento di Fisica, Università di Pisa, Largo B. Pontecorvo, 3, I-56127 Pisa, Italy
}%

\date{02 Oct 2023}

\begin{abstract}
The quantization of Yang-Mills  field theories requires the introduction of  a gauge fixing which leads to a violation of the Local Gauss Law described by the so-called Gauss operator. We discuss the local  quantizations  of   Yang-Mills theories in terms of the possible Gauss operators, which are argued to  have a more physical meaning than the gauge fixings.
  
We focus the attention on the local quantizations which leave the global gauge group  and a subgroup of local gauge transformations unbroken, as the Feynman quantization of quantum Electrodynamics, and show that in the non-abelian case such properties cannot be realized together with Lorentz covariance;  thus,  quite generally, one cannot have the structural properties of the Feynman quantization of  Quantum Electrodynamics. 

By relaxing the condition of Lorentz covariance, we obtain a classification of  Gauss operators, which satisfy  gauge covariant conservation laws and generate  non-trivial  residual subgroups of local gauge transformations.
\end{abstract}

\maketitle

\section{Introduction}
Given a Lie \textit{(global) gauge group} $G$, the correspondent \textit {local gauge group} $\G$ is conventionally obtained by replacing the constant group  parameters of $G$ by infinitely differentiable functions, which is convenient to choose of compact support in space and of slow (i.e. at most polynomial) increase in time\footnote{Regularity conditions on the gauge functions are needed, since the quantum fields are operator valued distributions, so that the transformations induced by   the action of $\G$   should have the property of multipliers   of the related test functions. The requirement of compact support in space guarantees the very useful property of space localization of the gauge transformations and it also provides the needed clear cut distinction between the global and the local gauge group. For the time dependence it is enough a slow increase in time; the condition of compact support in time would exclude the local gauge transformations in the temporal  gauge and it is not required. Technically, the group parameters of $\G$ are required to belong to    $ \D(\Rbf^3) \times \O_M(\Rbf) $).   Very problematic is the generalization of the local gauge transformations with the gauge functions replaced  by field operators, since it involves the singular (actually divergent) multiplication of operator valued distributions at the same point. For more detailed motivations and useful consequences of such choices see \cite{FS16}.}.

A classical gauge field theory is defined by a Lagrangian $\L_{inv}$ invariant under a local gauge group $\G$, with the fields (which enter in the definition of $\L_{inv}$) transforming as representations of $\G$. In order to have a well posed Cauchy problem, i.e. a deterministic time evolution of the fields, one must add a gauge fixing  $\LGF$, which need not  break  $\G$ to the identity (as implied by the functional integral argument), but only to the extent of allowing a deterministic field dynamics \cite{FS16} (for example, the Feynman gauge in electrodynamics leaves the unbroken residual group with parameters satisfying $\square\eps=0$). 

The introduction of a gauge fixing is crucial for a quantization of the field algebra since, according to the principles of quantum mechanics, the dynamics is described by a unitary operator $U(t) = e^{-i t H}$, and therefore  in order  to have \textit{quantum} fields the dynamics of the field operators must be deterministic.

\def \Fob  {\F_{obs}}
Thus, a quantization of a gauge field theory is defined  by the choice of a gauge fixing, different gauge fixings identifying different quantum field algebras $\F$, even if the resulting  representations of the observable field subalgebra $\F_{obs}$ are all equivalent; one of the main roles of the field algebra $\F$ is to allow, through its vacuum representation, the construction of the representations of $\Fob$ beyond its vacuum sector \cite{FS16}, namely  the expectations of the observables on the states with non-trivial  $G$ charges.

We  consider quantizations of Yang-Mills  field theories (with $G$ a compact Lie group) defined by  vacuum representations of  local field algebras $\F$,
in a vector  space $\V = \F \Psio$,  
 obtained by  \textit {gauge fixings which do not break the global gauge group $G$,  satisfy locality and leave a  residual local gauge subgroup $\G_r \subset \G$ unbroken }. 
\goodbreak

We briefly mention the motivations for such a choice.

Gauge fixings which break the global gauge group $G$ down to the identity (like the unitary gauge and the $\xi$ gauges)  
involve a mean field parameter which is difficult to justify beyond the perturbative approach and  it is at most  allowed for the discussion of the phenomenon of gauge symmetry breaking, as in the electroweak theory.

 A residual local gauge group $\G_r$ is very useful for exploiting the related Ward identities, a crucial tool for the proof of renormalization, as displayed by the prototypical case of  Feynman  gauge in Quantum Electrodynamics.   

Furthermore, the unbroken residual group allows us to  exploit its topology for codifying the interplay between gauge invariance and  other symmetries, in particular chiral symmetries. Indeed,  the representation of the topology of the residual group  yields   a non-perturbative  solution of the $U(1)$ problem in Quantum Chromodynamics, without relying on the semi-classical instanton approximation, which is not free of mathematical problems \cite{FS16}.   

The consideration of \textit{local} field algebras, i.e. of gauge fixings which do not violate locality, is motivated by the fact most of what is known on quantum field theories at the non-perturbative level relies on locality \cite{Haag}, which is  also  shared by the (perturbatively) renormalizable gauges.

Another ingredient of our analysis is the distinguished role of the Gauss operators, for the classification of the possible quantizations of Yang-Mills theories, since they have a more physical status than the gauge fixings, as argued  below.

In fact,  the presence of the gauge fixing invalidates the conclusion of the second Noether theorem, namely the validity of \textit{Local Gauss Law}
\be {\label{eq:1.1}J_\mu^a = \partial^\nu F^a_{\nu\,\mu},}\ee
 satisfied by the currents $J_\mu^a$, ($a = 1, ... n,  n = $ the dimension of $G$),
 which generate the infinitesimal transformations of $G$ on the field algebra $\F$:  
\be{\label{eq:1.2} \delta^a F = \limR\,i \, [ Q^a_R, \,F\,], \,\,\,\,\,\forall F \in \F, }\ee
where $Q^a_R$ is  a suitably regularized integral of $J^a_0$ over a sphere of radius  $R$.

Hence, instead of \cref{eq:1.1} one rather gets 
\be{\label{eq:1.3}J_\mu^a = \partial^\nu F^a_{\nu\,\mu} + G_\mu^a,}\ee
where $G_\mu^a$, henceforth called the \textit{Gauss operator}, depends on the choice of the gauge fixing.

The unbroken global group $G$ implies that $\partial^\mu  J^a_\mu = 0$ and therefore \cref{eq:1.3} and the antisymmetry  of $F_{\nu \, \mu}$ yield 
\be{\label{eq:1.4} \partial^\mu G^a_\mu = 0,}\ee 
i.e. \textit{the Gauss operator satisfies a continuity equation}. Thus, it plays the role of an \textquotedblleft external\textquotedblright{} conserved current, (a meaning not shared by the gauge fixing).

Moreover,   for any physical state $\Psi$ (typically obtained by a non-local construction through the vacuum representation  of $\F$, as for example in the Gupta-Bleuer quantization of QED),   the Local Gauss Law holds (for more details and proofs, see \cite{FS16}):
\be{\label{eq:1.5} (\Psi, (J_\mu^a - \partial^\nu F^a_{\nu\,\mu})\, \Psi) = (\Psi, G^a_\mu\, \Psi) = 0.}\ee
Actually, such a property qualifies the physical states (in Quantum Electrodynamics amounts to the validity of the Maxwell equations on the physical states), so that, by \cref{eq:1.5},   \textit {the Gauss operator provides  the characterization of the physical states}.

In addition,   by the locality of the field algebra it has been shown \cite{FS23} that in the Feynman quantization of QED and in general in the temporal gauge of Yang-Mills theories \textit{the Gauss operator generates the infinitesimal transformations of the residual local gauge group} in the given representation of $\F$, so that it  belongs to the commutant of the $\Fob$\, and therefore it classifies the representations of $\Fob$ contained in the representation space of $\F$. In the abelian case, the Gauss operator  even belongs  to $\Fob$ and therefore to its center.

The role and meaning  of the Gauss operator in such a construction of the physical states become more evident in connection with locality. 

In fact,  the Gauss law, \cref{eq:1.1},   implies  that the flux of $F^a_{i\,0}$ at infinity, briefly called \textit{Gauss charge} and denoted by  $Q^a_G$,  coincides with the total charge,  a suitable integral of  $J^a_0$, called the  \textit{current charge} and denoted by $Q^a_J$. Then,  since by locality the Gauss charge has vanishing expectations  on  local states,  so does the charge $Q^a_J$ and, as a consequence,  the global gauge group $G$  acts trivially in the representation space of a local field algebra $\F$. 
 
Hence, the role of the Gauss operator is to \textit{screen/compensate  the contribution  of the current charge to the flux of $F^a_{i\,0}$ at infinity} allowing  
for local states with non-zero charge and, consequently, for a non-trivial action of $G$ in the representation space of $\F$.

Furthermore, a clever choice of the Gauss operator may make more explicit the hyperbolic and elliptic  contents 
of \cref{eq:1.3}, which  in the abelian case is equivalent to: 
\be{\label{eq:1.6}\square  F_{\mu \,\nu} = \partial_\mu J_\nu - \partial_\nu J_\mu  - \partial_\mu G_\nu + \partial_\nu G_\mu ,\quad\partial^i F_{i \,0} = J_0 - G_0.}\ee

In fact, in the abelian case the choice of the temporal gauge $G_i = 0$, which  implies $\partial_0  G_0 =0$, leads to a standard hyperbolic  equation for
$F_{i\,j}$, 
\be{\label{eq:1.7}\square  F_{i \,j} = \partial_i J_j - \partial_j J_i,}\ee
without involving  the Gauss operator, which  only enters  in the \textquotedblleft elliptic\textquotedblright{} constraint involving  $F_{i\, 0}$ and $J_0$, consistent with the time evolution of $F_{i \, 0}$
\be{\label{eq:1.8}\partial^0 F_{0 \, i} = \partial^j F_{ i \, j} + J_i.}\ee
Thus, the role of $G_0$ is that of a screening  \textit {static charge density} which is seen by the local states but not by the \textit{non-local} physical charged states.

Furthermore, in the temporal gauge the Gauss operator satisfies a (gauge) covariant conservation law, namely $D^\mu G_\mu = 0$, in agreement with its interpretation as the current of a (fictitious) charge distribution.   

A similar role and meaning of the Gauss operator occurs also in the temporal gauge of the non-abelian case.

For these reasons, for the characterization of the possible (local) quantizations of a gauge field theory, the Gauss operator is better qualified than the gauge fixing, which does not have such properties. 

The aim of this note is to discuss  the  constraints on the Gauss operators in the non-abelian case (and therefore on the possible quantizations)  and point out the general obstruction for a quantization with the properties   similar to that of the Feynman gauge in Quantum Electrodynamics. 

We show that it is impossible to have a quantization of  non-abelian Yang-Mills field theories which satisfies locality, Lorentz covariance and leaves the global gauge group $G$ and a subgroup $\G_r \subset \G$ of local gauge transformations unbroken. This provides  a general argument for the recourse to BRST quantization in the case of  non-abelian Yang-Mills theories.
 
In addition, we  obtain a classification of the  Gauss operators which satisfy a (gauge) covariant conservation law. This property yields that   the gauge fixing and the corresponding Gauss operator minimally affect the gauge group  geometry of the Yang-Mills equations, being  required in particular if the current which generates the global group has to be  left unchanged by the gauge fixing.  

As a consequence, one has the desirable property that the action on the matter fields by the Gauss operator and by the topological operators  (namely, the operators which describe the topology of the residual group) is the same as that of the current in the absence of gauge fixing. 

 This pattern does not seem to be available in the BRST  quantization \cite{WE}: the invariance group of the Lagrangian is no longer the  group $\G$ of local gauge transformations but  rather the BRST group which involves field transformations with parameters which are field operators. 
As discussed above,  the mathematical control of  such  transformations, which involve singular multiplication of distributions at the same point, is rather problematic, beyond the perturbative approach. 

The BRST gauge fixing does not leave  a subgroup $\G_r$ of local gauge transformations unbroken and  the singular (distributional) properties of the fields which parametrize the BRST transformations preclude the possibility of considering their  topological properties. 

From this point of view, the   quantization of the temporal gauge proved more effective for a non-perturbative  analysis of non-abelian Yang-Mills theory, at the only expense of losing manifest Lorentz covariance. In exchange, the vacuum representation of the field algebra of the temporal gauge satisfies positivity and does not involve ghosts (so that the fields may be defined as   operators  in a Hilbert space), whereas the vacuum representation of the BRST field algebra is indefinite  and, in particular, the ghosts are not operators in a Hilbert  space. Needless to say, the identification or construction of the Hilbert space of physical states is much easier if the vacuum representation of the field algebra has a Hilbert space structure.
\goodbreak

\section{Gauss operators in  Yang-Mills gauge theories}
\label{secII}
For the reasons discussed above, in the following  we consider the possible  realization of a  Gauss operators $G_\mu$ which has   the following properties:

\begin{enumerate}[i)]
	\item it leaves the \textit{global gauge group $G$  unbroken},
	\item it leaves  \textit {a residual group $\G_r$ of local gauge transformations} unbroken; this  allows  for the exploitation of the corresponding Ward identities, and  in the non-abelian case the effect of the topology of $\G_r$ on the properties of the physical states,
	\item it defines  a \textit {local field},  in the representation space $\V$ of the local field algebra,
	\item it allows a simple \textit{characterization of the  physical states}  in terms of   the   subspace of the representation space of $\F$, or of a suitable completion of it, with vanishing expectation of $G_\mu$.\footnote{ For examples which realize such a structure see~\cite{FS16}.}  
\end{enumerate}
\goodbreak

\subsection{Group theoretical constraints on Gauss operators}

Since by ii) the Gauss operators under our consideration are related to residual subgroups of local gauge transformations, the characterization of such subgroups provides  group theoretical  constraints on the analysis of possible Gauss operators. 

As mentioned above, the local gauge group $\G$ is  parametrized by the  set of infinitely differentiable functions of compact support in space, $\eps^a(x)$, $a = 1, ...n$, which form not only a vector space $\V$ with the sum defined by 
\[  (\eps_1 + \eps_2)^a(x) \eqq \eps_1^a(x)  + \eps_2^a(x),\]
but also an algebra $\E$ with product 
\[ (\eps_1 \wedge \eps_2)^a(x) \eqq i f^a_{b\, c} \eps_1^b(x) \, \eps_2^c(x).\]
The abelian subgroups $\G_i$ of $\G$, stable under $G$ by i), are identified by subspaces $\V_i$ of $\V$, whereas the non-abelian subgroups are identified by sub-algebras $\E_i$ of  $\E$, since in the non-abelian case the group composition law requires stability under the above product (see the Zassenhaus formula  below).

The stability of $\G_i$ under the global group $G$ requires that if $\eps^a(x) T^a$ belongs to the Lie algebra of $\G_i$ so does $ [\,\eps^b\,T^b, \eps^a(x) T^a\, ], \forall b, $ i.e. if $\eps^a$ belongs to the parameter space of $\G_i$ so does $\eps^c(x) =  f^c_{a \, b} \eps^a(x), \,\forall b$.

The above properties must be satisfied by the local gauge subgroups  related to the Gauss operators and therefore provide a first classification of them. 

Another restriction comes from the condition that the Gauss operators  admit a derivation from a Lagrangian.

The invariance of the gauge fixing  under the residual group $\G_r$, with gauge parameters $\eps_r$ is codified by the following equations in terms of the gauge fixing $\LGF$: 
\begin{gather}
	\label{eq:2.1}
	L_a^\mu \,\partial_\mu \eps^a_r  \eqq \\ \nonumber\left \{ \frac{\delta \LGF}{\delta A^a_\mu} + i T^a_{b c} \left [ \frac{\delta \LGF}{\delta\, \partial_\mu A^b_\nu} A^c_\nu +  \frac{\delta \LGF}{\delta \,\partial_\mu B^b} B^c \right]  \right \}\, \partial_\mu \eps_r^a = 0 , \\
	\label{eq:2.2}
	L_a^{\mu\, \nu} \,\partial_\mu \partial_\nu \eps^a_r \eqq\frac{\delta \LGF}{\delta\, \partial_\mu A^a_\nu} \partial_\mu \partial_\nu \eps^a_r =0,
\end{gather}
where for simplicity, we omitted the presence  of matter fields and considered gauge fixings $\LGF$ which are functions of $A^a_\mu$ and of fields $B^a$, of the Nakanishi-Lautrup type,  which transform according to irreducible adjoint representations of $G$,  taken to be a  simple group.  

In conclusion, in the case of an abelian global group $G$ the possible residual groups, equivalently the possible Gauss operators, are identified by $G$ invariant  subspaces of $\V$ with gauge parameters $\eps_r(x)$ satisfying~\cref{eq:2.1,eq:2.2}.

For non-abelian $G$ the characterization is provided by $G$ invariant sub-algebras of $\E$ with gauge parameters $\eps_r(x)$ satisfying~\cref{eq:2.1,eq:2.2}.  

Explicit cases are discussed below.
\goodbreak

\subsection{Feynman Gauss operator}
\label{secIIB}
A distinguished choice of the Gauss operator is that of the Feynman gauge in Quantum Electrodynamics, which preserves locality and covariance: $G_\mu = \partial_\mu L$, with $L$ a massless  longitudinal field. 

In addition to the properties i)-iv),  the Feynman Gauss operator in QED satisfies 
\begin{enumerate}[i)]
	\setcounter{enumi}{4}
	\item it is \textit{Lorentz covariant}, allowing for a Lorentz covariant  field algebra,
	\item it provides a simple \textit{subsidiary condition}, which linearizes the constraint \cref{eq:1.5} for the identification of the physical states,  
	\item it \textit{does not require ghost fields}, since it  involves  only the vector potential and possibly the Nakanishi-Lautrup  field $B$.  
\end{enumerate}

The technical advantages of such a choice are well known, in particular for the control of renormalization,  the explicit construction of the physical charged states, etc. (see e.g.~\cite{FS16}).
It is then natural to ask whether  a Gauss operator  with such properties  may be obtained  also in the non-abelian case.

A negative answer  has been presented  in the literature and it is part of the standard  textbook presentation of the need for BRST quantization.  However, such a negative conclusion has been reached  for the choice of a gauge fixing which is a
literal transcription of the Feynman choice, namely  for the gauge fixing $\ume \xi (\partial^\mu A^a_\mu)^2$ or for its  Nakanishi-Lautrup variations, e.g. $ B^a \partial^\mu A^a_\mu +  \ume \xi (\partial^\mu A^a_\mu)^2$. 

One may reasonably  ask whether in  the non-abelian case a more clever choice of the gauge fixing may work. 
Quite generally this leads to investigating the possible Lorentz covariant solutions of the equation $\partial^\mu G_\mu = 0.$
\goodbreak

For simplicity, we discuss the problem in the absence of matter fields and consider gauge fixings $\LGF$ which are functions of $A^a_\mu$ and of fields $B^a$, of the Nakanishi-Lautrup type,  which transform according to irreducible adjoint representations of $G$, for simplicity taken to be a  simple group.  
Then,  the invariance of the gauge fixing $\L_{GF}$  
under the global gauge group $G$ requires
\begin{align}
   T^a_{b \, c}\,\left [ \frac{\delta \LGF}{ \delta A^b_\mu} A^c_\mu  + \frac{\delta \LGF} {\delta B^b} B^c  +\right.\\
   \left.+\frac{\delta \LGF}{ \delta \partial_\nu A^b_\mu} \partial_\nu A^c_\mu  + \frac{\delta \LGF} {\delta \partial_\nu B^b} \partial_\nu B^c \right ] = 0
\end{align}

The additional restriction of Lorentz covariance plays a crucial role, requiring the Lorentz covariance of the subspaces of gauge parameters in the abelian case and of the sub-algebras of $\E$ in the non-abelian case.
 Then,  the group parameters  of $\G_r$ should define a vector space stable under $G$ and under Lorentz transformations. Now, since $G$ does not have invariant subgroups, \cref{eq:2.1} cannot provide restrictions on the index $a$, and similarly there is no linear subspace invariant under the Lorentz transformations to be spanned by the four vectors $\partial_\mu \eps^a$.  

Then, in the absence of restrictions on the group parameters, in order to satisfy~\cref{eq:2.1} the expression in curly brackets should vanish  
\be {\label{eq:2.3}\left \{ \frac{\delta \LGF}{\delta A^a_\mu} +i T^a_{b c} \left [ \frac{\delta \LGF}{\delta\, \partial_\mu A^b_\nu} A^c_\nu +  \frac{\delta \LGF}{\delta \,\partial_\mu B^b} B^c \right]  \right \} = 0.}\ee

It is instructive to check that~\cref{eq:2.1} provides restrictions on the  gauge parameter $\eps_r$  if Lorentz covariance is not required, as it happens for the temporal gauge. In fact, in this case  the term in square brackets  and $\delta \LGF/\delta A^a_i,  i=1, 2, 3$ vanish, (see below),  and~\cref{eq:2.1} gives  the restriction $ \partial_0 \eps_r = 0$, i.e. the residual group of the  temporal gauge. 

It remains to exploit~\cref{eq:2.2}, which (assuming Lorentz invariance) yields the following characterization  
\be{\label{eq:2.4}\frac{\delta \LGF}{\delta\, \partial_\mu A^a_\nu} + \frac{\delta \LGF}{\delta\, \partial_\nu A^a_\mu} 
= g_{\mu \, \nu} L^a,\qquad \square \,\eps^a_r(x) = 0, }\ee
with $L^a \neq 0, $ a scalar function of the fields. Clearly, if  $L^a = 0$, the condition $\square \,\eps^a = 0$ is not required, but then the absence of conditions on the $\eps^a$ would imply $\G_r = \G$, i.e. no effective gauge fixing. 

Hence, the possible residual subgroup should be characterized by gauge parameters $\eps^a(x) $ satisfying the (free)  wave equation. Indeed, as we shall see, this works in the abelian case, but not in the non-abelian case.\goodbreak

In fact, in the non-abelian case there cannot be a residual group $\G_r$ characterized by gauge parameters $\eps_r$  satisfying $\square \eps_r = 0$, since this condition  keeps being satisfied 
by linear combination of such functions (vector space structure), but not by their products in conflict with the group composition rule.

This is  displayed by 
 the Zassenhaus formula,~\cite{MA, WI}, which is known to converge for small values of the parameters; in fact, it gives
\[ e^{i \eps^a_1(x)  t^a }\, e^{ i \eps^b_2(x) t^b}  = e^{i  \eps^a_1(x)  t^a  + i \eps^b_2(x) t^b} \, e^{ -\frac{1}{2}\eps^a_1(x)  \eps^b_2(x) [t^a, \,t^b]} ...\]
The first term on the right hand side is the exponential of an element of the Lie algebra, and therefore an element of the group, but the other factors, which do not vanish in the non-abelian case, do not define elements of a gauge group characterized by the above gauge parameters, because the product of gauge functions which satisfy  the wave equation does not satisfy the wave equation. 

In conclusion, quite generally \textit{in the non-abelian case it is impossible to have a covariant Gauss operator satisfying i)-ii), 
as a function of $A^a_\mu$ and possible Nakanishi-Lautrup fields}; hence the properties of the Feynman Gauss operator are in general  excluded. Lorentz covariance can be rescued by relaxing the condition of a non-trivial residual gauge group, as in BRST quantization.
\goodbreak

\subsection{Gauss operator and covariant conservation law}
\label{secIIC}
By the defining \cref{eq:1.3}, $\partial^\mu G_\mu= 0$, where $J^a_\mu$ is the current which generates the unbroken global gauge group $G$ thanks to  Noether theorem.

Now, in the absence of a gauge fixing, the Gauss law may also be written as 
\be{\label{eq:2.5} D^\nu F^a_{\nu \,\mu} = j^a_\mu(\psi), }\ee
where $ D^\mu $ denotes the covariant derivative and  $j^a_\mu(\psi)$ is the contribution to the conserved current by the matter fields (the so-called matter current). Both terms of this equation   satisfy a covariant conservation law
\be{\label{eq:2.6} D^\mu D^\nu F^a_{\nu \,\mu} = D^\mu j^a_\mu(\psi) = 0.}\ee
One may then ask whether such covariant conservation laws, strictly related to the gauge group geometry, keep holding also in the presence of  a gauge fixing. Thus we look for Gauss operators which satisfy a covariant conservation law (besides the ordinary conservation law), i.e.
\be{\label{eq:2.7}D^\mu G^a_\mu  = 0.}\ee

We allow for gauge fixings which are functions of $A^a_\mu$ and of auxiliary fields of the Nakanishi-Lautrup type $B^a_i, i = 1,...k$, which transform as the adjoint representation of $G$, and possibly  $G$ invariant fields $B_i$.

We shall see that if 

\noindent a) the gauge fixing does not depend on the derivatives of  $A^a_\mu$, and
\vspace{2pt} 

\noindent b) the current 
\[ j^a_\mu(B)  =  \frac{\delta \L}{\delta \partial_\mu B^b_i}  \delta^a  B^b_i     =   \frac{\delta \LGF}{\delta \partial_\mu B^b_i} \delta^a B^b_i \]
satisfies a covariant conservation law (in particular $j^a_\mu(B) = 0$ if $\LGF$ does not depend on the derivatives of the $B^a_i$) 
\vspace{2pt}

\noindent then, the corresponding Gauss operator satisfies a covariant conservation law.
Indeed  we shall see that, contrary to what implicitly argued in the literature, there are solutions of~\cref{eq:2.7} and of the continuity equation.     

\def \Lo  {\L_0}

We denote by $\Lo$ the gauge invariant Lagrangian , so that $\L = \Lo + \LGF$.
By  Noether theorem, the invariance of $\L$ under the global gauge group $G$ determines the conserved current $J^a_\mu$ which generates the transformations of $G$: 

\begin{widetext}
\begin{align}
	\label{eq:2.8}
	J^a_\mu = i\,   T^a_{b c} \left [ \frac{\delta \L}{\delta\, \partial_\mu A^b_\nu} A^c_\nu +  \frac{\delta \L}{\delta \,\partial_\mu B^b_i} B^c_i \right] + j^a_\mu(\psi)=\\
	\label{eq:2.9}
	= i\,   T^a_{b c} \left [ \frac{\delta \Lo}{\delta\, \partial_\mu A^b_\nu} A^c_\nu +  \frac{\delta \LGF}{\delta \,\partial_\mu B^b_i} B^c_i \right]  + j^a_\mu(\psi) = 
	j^a_\mu(A) + j^a_\mu(B_i) + j^a_\mu(\psi),
\end{align}
\end{widetext}

where $j^a_\mu(\psi)$ denotes the contribution of the matter fields, the so-called matter current,  and $j^a_\mu(A) = - i  T^a_{b \, c} \,(A^b)^\nu  F^c_{\nu \mu}$.

The second Noether theorem applied to $\L - \LGF$ provides information on the effect  of $\LGF$ in yielding~\cref{eq:1.3}. 

The coefficient of $\partial_\mu \eps$ in the infinitesimal variation of $\Lo =\L - \LGF$ must vanish and one gets  
\begin{align}
	\nonumber
	J^a_\mu + \frac{\delta \L}{ \delta A^a_\mu} - \frac{\delta \LGF}{ \delta A^a_\mu}  - j^a_\mu(B_i) =\\  J^a_\mu + \partial^\nu \frac{\delta \Lo} {\delta \partial _\nu A^a_\mu} - \frac{\delta \LGF}{ \delta A^a_\mu} -j^a_\mu(B_i) = \\
	\label{eq:2.10}
	 = J^a_\mu  + \partial^\nu F^a_{\mu\, \nu}  - \frac{\delta \LGF}{ \delta A^a_\mu} - j^a_\mu(B_i) = 0 ,
\end{align}

where we have used that 

\noindent 1) $\LGF$ does not depend on the derivatives of $A^a_\mu$,

\noindent 2) $\Lo$ does not depend on the auxiliary fields $B^a_i$, so that  the derivatives of $\LGF$ with respect to the fields $B^a_i$ and to their derivatives may be replaced by the corresponding derivatives of  $\L$, allowing to use the   equations of motion for the fields $B^a_i$.

Hence, the  Gauss operator  is given by 
\be{\label{eq:2.11}G^a_\mu =  \frac{\delta \LGF}{ \delta A^a_\mu} + j^a_\mu(B_i).}\ee
and~\cref{eq:2.10} gives 
\be{\label{eq:2.12} D^\nu F^a_{\nu \mu} = j^a_\mu(\psi)  + j^a_\mu(B) - G^a_\mu.}\ee
Now, the matter current satisfies a covariant conservation law and,  by assumption so does  also $j^a_\mu(B_i)$:
\[ D^\mu j^a_\mu(B_i) = 0.\]

Then, since $ D^\nu D^\mu F^a_{\nu \, \mu} = 0$,~\cref{eq:2.10} gives 
\be {\label{eq:2.13} D^\mu   G^a_\mu  = 0.}\ee

\subsection{Solutions of the  equations  $\partial^\mu G^a_\mu = 0,\;    D^\mu   G^a_\mu  = 0 $}
\label{secIID}
Given a solution $G^a_\mu$ of the conservation law $\partial^\mu G^a_\mu = 0 $, the additional covariant conservation law  requires 
\be{\label{eq:2.14} T^a_{b \,c } \, (A^b)^\mu G^c_\mu = 0.}\ee 

\vspace{1mm}

The vanishing of the left hand side may be due  either to the summation over  the Lorentz indices or to the summation over the $G$-group  indices $b, c$ (for each Lorentz index $\mu$).

We have  already  shown  that in the non-abelian case the equation $\partial^\mu G^a_\mu = 0 $ does not have  Lorentz  covariant solutions fulfilling our requirements. One can  see that in the non-abelian case also~\cref{eq:2.14} does not admit Lorentz covariant solutions. 

In fact, the invariance of the solutions under the corresponding residual group requires 
\[  T^a_{b \,c } \,\partial^\mu \eps^b_r(x)\, G^c_\mu = 0. \]
Since the parameters $\partial^\mu \eps_r(x)$ span a Lorentz covariant vector  subspace  and no  such subspace exits except the whole Lorentz covariant vector space,   
\be\label{eq:2.15}(v^b)^\mu\, T^a_{b \,c } \, G^c_\mu = 0, \ee
for any Lorentz vector $v^b_\mu$. Therefore  $G^c_\mu = 0$ and $\G_r = \text{\usefont{U}{bbold}{m}{n}1}$.
 
Thus, one must look for solutions which violate Lorentz covariance, still keeping the conditions i)-iv).

Now, the covariance of~\cref{eq:1.3} under the unbroken global group requires that  the Gauss operator must transform according to the adjoint representation. 

 Therefore,  from a group theoretical point of view, in terms of the  (irreducible) adjoint representation of $SU(N)$, \cref{eq:2.13} may be written in the form of  scalar  product
 \[(G, T^a A) = 0,\]
 where $A$ and $G$ are  vectors of the (irreducible) adjoint representation.

\noindent Now, in an irreducible representation given  any vector by applying    the generators $T^a$ and the identity one reaches any other vector; therefore the above equation requires $G \sim A$.

Thus, the solutions of~\cref{eq:2.13} are $G^a_\mu = A^a_\mu B$, with $B$ an operator invariant under $G$ transformations. This is easily obtained in the case $G = SU(2)$, since then~\cref{eq:2.13} requires the vanishing of the vector product $ \bf{A}_\mu \wedge \bf{G}_\mu$.

A very relevant example is provided by the \textit{Gauss operator of the temporal gauge}, characterized by  $G^a_i = 0, \, \partial_0 G^a_0 = 0$.

It is uniquely selected by the condition of rotational covariance of $G^a_\mu$. In this case if $v^b_\mu = (v^b_i, v^b_0)$  is an allowed  vector, so are the space-rotated ones  $Rv^b_\mu = ((Rv^b)_i, v^b_0)$ as well as  $ Rv^b_\mu - v^b_\mu = ((Rv^b)_i - v^b_i, 0)$ for any rotation $R$. The set of such vectors spans the whole three-dimensional space and therefore \cref{eq:2.15} implies $G^c_ i = 0$.

Its explicit form is obtainable by the gauge fixing term $\LGF = A^a_0\, B^a$, with $B^a$ a Nakanishi-Lautrup field which plays the role of a Lagrange multiplier leading to $A^a_0 = 0$. 

Then, one has  
\be{\label{eq:2.16}G^a_\mu = \delta \LGF / \delta A^a_\mu = \delta_{\mu, 0}\, B^a, }\ee
 and~\cref{eq:2.13} is obviously satisfied since $G^a_i = 0$ and $A^a_0 = 0$.
\goodbreak

 In this case, the residual local  covariance group of~\cref{eq:1.3} is given by time independent gauge parameters,  which characterize the \textit{temporal gauge}. Indeed,  the constraint of~\cref{eq:2.1} is satisfied by unrestricted $ \partial_i \eps^a$, which requires    $\delta \LGF/\delta A^a_i = 0$, and by $\partial_0 \eps^a = 0$, which allows for $\delta \LGF/\delta A^a_0 \neq 0$.
  
This realization of the temporal gauge is particularly convenient because it directly  provides the violation of the Local Gauss Law in terms of the time independent Nakanishi-Lautrup local field $B^a$. It is one of the basic fields which generate the local field algebra and it does not arise as a composite field with the related problem of point splitting regularization; thus, its algebraic properties are derivable from canonical quantization.

A similar pattern is realized by the \textit{axial gauge} or by variations of it, like e.g. the $n^\mu A^a_\mu = 0 $ gauges, with $n_\mu$ timelike, spacelike or lightlike. In this case, rotational covariance is replaced by the covariance under the stability group of $n_\mu$. 

\section{Conclusion}
We have shown that it is impossible to quantize Yang-Mills gauge theories with the properties of the Feynman quantization of Quantum Electrodynamics, namely with a Gauss operator which is Lorentz covariant  and preserves a residual subgroup of local gauge transformations. If the covariance condition is relaxed, one may  obtain a classification of the  Gauss operators, which satisfy (gauge) covariant conservation laws and generate non-trivial  subgroups of local gauge transformations.   

It is worthwhile to remark that most of the non-perturbative results on gauge field theories (see the discussion in \cite{FS16}). have been obtained by using such a class of Gauss operators, thanks to their properties of a) preserving locality, b) leaving a residual local gauge group unbroken (helpful for renormalization), whose topology plays a crucial role in solving the $U(1)$ problem and  c) satisfying a covariant  conservation law, which  allows for a simple control of its action on the matter fields.  Moreover, the vacuum representation of the correspondent field algebra satisfies positivity and therefore has a Hilbert space structure.

Thus, on one side our results provide a general argument for justifying  the recourse to Becchi-Rouet-Stora-Tyutin (BRST) quantization  in  non-abelian Yang-Mills theories in order to have manifest Lorentz covariance and locality, even if the residual group is not a subgroup of local gauge transformations and involves transformations with ghost fields as  group parameters, violating positivity. On the other side it adds further support for the distinguished role of the temporal gauge, which proved very useful for  a non-perturbative approach and results in non-abelian gauge theories.

\section{Acknowledgments}
B.B. is partially supported by INFN Iniziativa Specifica TPPC.

\end{document}